\documentclass[twocolumn,amsmath,amssymb,amsfonts,superscriptaddress,floatfix,showpacs,prd,aps,nofootinbib,preprintnumbers]{revtex4}
\newcommand{\del}{\partial}
\newcommand{\Tr}{\mathrm{Tr}}

\newcommand{\vev}[1]{\langle{#1}\rangle}

\usepackage{graphicx,color,mathrsfs,flafter
}

\begin{document}
\title{ Role of mesonic fluctuations in
  the Polyakov loop extended quark-meson model at imaginary chemical potential}

\author{Kenji Morita}
\affiliation{Yukawa Institute for Theoretical Physics, Kyoto University,
Kyoto 606-8502, Japan}
\affiliation{%
ExtreMe Matter Institute EMMI, GSI, D-64291 Darmstadt, Germany}

\author{Vladimir Skokov} \email[E-Mail:]{V.Skokov@gsi.de} \affiliation{%
  GSI Helmholtzzentrum f\"ur Schwerionenforschung, D-64291 Darmstadt,
  Germany}

\author{Bengt Friman} \affiliation{%
  GSI Helmholtzzentrum f\"ur Schwerionenforschung, D-64291 Darmstadt,
  Germany}

\author{Krzysztof Redlich} \affiliation{%
  Institute of Theoretical Physics, University of Wroclaw, PL--50204
  Wroc\l aw, Poland}
\affiliation{%
ExtreMe Matter Institute EMMI, GSI, D-64291 Darmstadt, Germany}

\preprint{YITP-11-71}

\begin{abstract}
 We explore the thermodynamics and phase structure of the Polyakov loop-extended
 two flavor chiral quark--meson (PQM) model beyond the mean-field
 approximation at imaginary chemical potential.	Our approach is based
 on the functional renormalization group (FRG) method. At finite
 temperature and imaginary chemical potential, we solve the renormalization
 group flow equation for a scale-dependent thermodynamic potential in
 the presence of the gluonic background field.	We determine behavior
 of order parameters of the PQM model in the FRG approach and compute the phase diagram.
 We compare our FRG results with that obtained in the mean-field
 approximation at imaginary chemical potential.
\end{abstract}

\maketitle

\section{Introduction}
Thermodynamic properties of strongly interacting matter at nonzero
baryon density and at finite  temperature have been explored numerically
within the Lattice Quantum Chromodynamics (LQCD)
\cite{fl2,L_lgt1,L_lgt2,MNNT_PTPreview,Kaczmarek2011,fl1}.
The LQCD results show that QCD at physical quark masses and at vanishing
baryon density exhibits restoration of the chiral symmetry and
deconfinement at $T\simeq 160$ MeV  \cite{WB1,fodor,HISQ}. The crossover
nature of the transition
at zero baryon density makes the variation of the order parameters quite
smooth and leads to the broader width of the pseudocritical temperature~\cite{fodor}.

Unfortunately, the thermodynamics of strongly interacting matter at
large baryon densities 
is presently not
accessible in the first principle LQCD calculations, because  of a complex structure of 
fermion determinant.
Several methods circumventing this, so-called, sign-problem have been introduced.
{ However, they are  limited to  small values of chemical potentials or small lattice sizes~\cite{deForcrand09}.
Thus, LQCD calculations have not been able yet to
address a nature of a critical behavior at finite density on
sufficiently large lattices  and at small quark masses.}

Phenomenological models and effective theories offer a viable
framework for exploratory studies.
The properties of low-energy hadrons as well as the nature of the
chiral phase transition at finite temperature and density have been
studied intensively in such effective models ~\cite{Gocksch,Hatsuda-Kunihiro,Buballa:review,Meisinger,Fukushima,Mocsy,PNJL,CS,DLS,Megias,IK,Fukushima:strong,Schaefer:PQM,sasaki,Abuki08,Kahara2010,Palhares2010,EPNJL,PDSE,Aarts}.
{ Recently,  the physics of color confinement and its relation to the chiral
symmetry breaking have been also addressed in such a framework.
Particularly successful, are models obtained   
by extending the chiral Lagrangians, such as
the Nambu--Jona--Lasinio or the quark--meson, by introducing  a coupling
of quarks to uniform temporal background gauge fields,  the Polyakov loop ~\cite{Fukushima,Schaefer:PQM}.
Though, these models have been formulated so as to incorporate essential features of
QCD, nevertheless, properties of phase transitions at large baryon densities strongly
depend on their  parameterizations~\cite{Stephanov:2007fk,Ratti:2007jf,CS,Schaefer:PQM,Abuki08,kap}.}

{ The imaginary chemical potential formulation provides complementary constraints on the
phenomenological models. In addition, at imaginary $\mu$,  the phase
diagram can be directly computed in LQCD by virtue of the absence of the
sign problem. 
Within LQCD, these  results   have  been used to extrapolate the phase boundary from an 
imaginary to a real chemical potential \cite{deF-P,deForcrand:2007,D'Elia-Lombardo,Wu,Cea2010,Nagata}. Such a strategy was  found to be successful 
in two-color QCD \cite{Papa}, resummed}
perturbation theory \cite{Hart_PLB2001}, and quasi-particle models
\cite{Bluhm2008}. 

Another interesting issue, that can be addressed at imaginary chemical potential, is the order of the
transition at Roberge-Weiss (RW) endpoint, which  might dictate the
property of the transition at vanishing and finite real chemical potential.
Recently,  it was shown, within lattice simulations,  that 
the critical behavior at the RW endpoint has a non-trivial dependence on
the quark masses for both, two-\cite{Bonati_RW} and three-flavor QCD
\cite{deForcrand_RW}.

{ The formulation of the theory at imaginary $\mu$ also opens the  possibility to 
obtain the canonical partition function from LQCD by means of the Fourier
transformation of the grand canonical partition function  \cite{Alford,deForcrand_canonical,Ejiri,Li}.
 Therefore, studies of thermodynamics at  imaginary chemical potential
are important to explore  thermodynamics of the QCD at finite density and its phase diagram \cite{pawlowski}}.

The Polyakov loop extended Nambu--Jona--Lasinio
(PNJL) \cite{PNJL} and quark--meson (PQM) \cite{Schaefer:PQM} models
reproduce essential features of the QCD thermodynamics already in the
mean-field approximation at real and at imaginary chemical
potential~\cite{Kyushu_PNJL,Kenji:2011}.
Especially, the Roberge-Weiss (RW) periodicity
$Z(\mu_I/T)=Z(\mu_I/T+2\pi k/N_c)$
and the associated $Z(N_c)$ transition at high temperature \cite{RW} can be described
{ in these models.}

In our previous studies \cite{Kenji:2011}, we have discussed the phase
{ structure of the PNJL model at imaginary chemical potential.}  While the
statistical confinement feature of this  model naturally provides
characteristic properties of the order parameters,  the chiral condensate and
the Polyakov loop, {  its thermodynamics, however, is  strongly influenced} by  the choice of the
Polyakov loop potential. { Furthermore, to reproduce  within the mean field approximation the  LQCD results in terms of such  effective models,  several amendments seem to be necessary}
\cite{Kyushu_PNJL2,EPNJL,Kashiwa_nonlocalPNJL,Pagura}.

To correctly account for the critical behavior and scaling properties
near the chiral phase transition, it is necessary  to go beyond the
mean-field approximation and include fluctuations and
non-perturbative dynamics. This can be achieved e.g. by using methods
based on the functional renormalization group
(FRG)~\cite{Wetterich,Morris,Ellwanger,Berges:review,Schaefer:2006ds,SFR}.

In the following, we consider the PQM model at imaginary chemical
potential and study its critical properties and phase diagram within FRG
approach. We formulate and solve the suitably truncated FRG flow equation
for fluctuations of the meson fields in the presence of the Polyakov loop which is  treated
as a background field on the mean-field level. We extend previous
studies~\cite{Skokov:2010wb} to imaginary chemical potential and
explore influence of fluctuations on the chiral and deconfinement
order parameters.
 We compare our FRG results at imaginary chemical potential with that
 obtained in the mean field  approximation. We show, that there is an
 essential modification of thermodynamics at imaginary chemical
 potential  owing to quantum mesonic fluctuations.

The FRG approach was previously
applied to study the phase structure of two-flavor QCD in the chiral
limit at imaginary chemical potential~\cite{pawlowski}. 
In the current paper, we perform calculations within the PQM model to
study the role of the mesonic fluctuations on the structure of the phase
diagram at imaginary chemical potential by comparing results obtained
within the FRG approach and undeer the mean-field approximation. 


In the next section, we  introduce  the PQM model
and the  implementation of the FRG method at imaginary $\mu$. We  present our  results in
Sec.~\ref{sec:thermo}. Section \ref{sec:concl} is devoted to the
summary.

\section{The Polyakov-quark-meson model}\label{sec:pqm}

The quark--meson model is an effective realization of the low--energy
sector of QCD, which incorporates chiral symmetry. Because the
local color $SU_c(N)$ invariance of QCD is replaced by a global
symmetry, the model does not describe confinement. Nevertheless, by
introducing a coupling of the quarks to a uniform temporal color gauge
field, represented by the Polyakov loop, some  confinement properties 
can be effectively included  ~\cite{Fukushima,Fukushima:strong, Schaefer:PQM}.

The Lagrangian of the PQM model reads \cite{Schaefer:PQM}
\begin{eqnarray}\label{eq:pqm_lagrangian}
  {\cal L} &=& \bar{q} \, \left[i\gamma^\nu {D}_\nu  - g (\sigma + i \gamma_5
    \vec \tau \cdot \vec \pi )\right]\,q
  +\frac 1 2 (\partial_\nu \sigma)^2+ \frac{ 1}{2}
  (\partial_\nu \vec \pi)^2
  \nonumber \\
  && \qquad - U(\sigma, \vec \pi )  -{\cal U}(\Phi,\Phi^{*})\ .
\end{eqnarray}
The coupling between the effective gluon field and quarks is
implemented through the covariant derivative
\begin{equation}
  D_{\nu}=\del_{\nu}-iA_{\nu},
\end{equation}
where $A_\nu=g\,A_\nu^a\,\lambda^a/2$. The spatial components of the
gluon field are neglected, i.e. $A_{\nu}=\delta_{\nu0}A_0$.  Moreover,
${\cal U}(\Phi,\Phi^{*})$ is the effective potential for the gluon
field expressed in terms of the thermal expectation values of the
color trace of the Polyakov loop and its conjugate
\begin{equation}
  \Phi=\frac{1}{N_c}\vev{\Tr_c L(\vec{x})},\quad \Phi^{*}=\frac{1}{N_c}\vev{\Tr_c
    L^{\dagger}(\vec{x})},
\end{equation}
with
\begin{eqnarray}
  L(\vec x)={\mathcal P} \exp \left[ i \int_0^\beta d\tau A_4(\vec x , \tau)
  \right]\,,
\end{eqnarray}
where ${\mathcal P}$ stands for the path ordering, $\beta=1/T$ and
$A_4=i\,A_0$.  In the $O(4)$ representation, the meson field is
introduced as $\phi_m=(\sigma,\vec{\pi})$ and  the corresponding
$SU(2)_L\otimes SU(2)_R$ chiral representation is defined by
$\sigma+i\vec{\tau}\cdot\vec{\pi}\gamma_5$.

The purely mesonic potential of the model $U(\sigma,\vec{\pi})$, is
defined as
\begin{equation}
  U(\sigma,\vec{\pi})=\frac{\lambda}{4}\left(\sigma^2+\vec{\pi}
    ^2-v^2\right)^2-c\sigma,\label{eq:mesonic_potential}
\end{equation}
while the effective potential of the gluon field is parametrized in
such a way as to preserve the $Z(3)$ invariance,
\begin{equation}
  \frac{{\cal U}(\Phi,\Phi^{*})}{T^4}=
  -\frac{b_2(T)}{2}\Phi^{*}\Phi
  -\frac{b_3}{6}(\Phi^3 + \Phi^{*3})
  +\frac{b_4}{4}(\Phi^{*}\Phi)^2\,\label{eff_potential}.
\end{equation}
The parameters,
\begin{eqnarray}
  \hspace{-4ex}
  b_2(T) &=& a_0  + a_1 \left(\frac{T_0}{T}\right) + a_2
  \left(\frac{T_0}{T}\right)^2 + a_3 \left(\frac{T_0}{T}\right)^3\,
\end{eqnarray}
with $a_0 = 6.75$, $a_1 = -1.95$, $a_2 = 2.625$, $a_3 = -7.44$, $b_3 =
0.75$,  $b_4 = 7.5$ and $T_{0}=270$ MeV were chosen to reproduce the equation of state
of the pure $SU_c(3)$ lattice gauge theory.  When the
coupling to the quark degrees of freedom are neglected, the potential~(\ref{eff_potential}) yields a
first-order deconfinement phase transition at
$T_0$.  

Alternative parametrization of the Polyakov loop potential, see
e.g. Ref.~\cite{Ratti:2007jf}, may provide a better fit of the  lattice
results by the PQM model. However, we follow our previous
studies of the PQM model formulated at the real chemical potential, and
apply the potential~\eqref{eff_potential} in the model calculations at
imaginary $\mu$. The main conclusions on the
influence of the mesonic fluctuations on the phase structure will not
change if different parametrization of the potential is considered.

\subsection{The FRG method in the PQM model}\label{sec:rg}

In order to account for mesonic  fluctuations in the PQM
model, we employ a scheme based on the functional renormalization group (FRG).
This scheme involves an infrared regularization of the fluctuations at a sliding momentum scale
$k$, resulting in a scale-dependent
effective action $\Gamma_k$, the so-called effective average action~\cite{Wetterich, Morris, Ellwanger, Berges:review}.
We treat the Polyakov loop
as a background field, which is introduced self-consistently on the
mean-field level while fluctuations of  the quark and meson fields  are accounted for by solving the FRG flow equations.

\begin{widetext}
  Following our previous work~\cite{Skokov:2010wb}, we formulate the
  flow equation for the scale-dependent grand canonical potential density $\Omega_{k}=T\Gamma_{k}/V$ for
  the quark and meson subsystems at finite temperature $T$ and imaginary chemical potential $\mu=i \theta T$ as follows
  \begin{eqnarray}\label{eq:frg_flow}
    \del_k \Omega_k(\Phi, \Phi^*; T,\theta)&=&\frac{k^4}{12\pi^2}
    \left\{ \frac{3}{E_\pi} \Bigg[ 1+2n_B(E_\pi;T)\Bigg]
      +\frac{1}{E_\sigma} \Bigg[ 1+2n_B(E_\sigma;T)
      \Bigg]   \right. \\ \nonumber && \left. -\frac{4 N_c N_f}{E_q} \Bigg[ 1-
      N(\Phi,\Phi^*;T,\theta)-\bar{N}(\Phi,\Phi^*;T,\theta)\Bigg] \right\}.
  \end{eqnarray}
  Here $n_B(E_{\pi,\sigma};T)$ is the bosonic distribution function
  \begin{equation*}
    n_B(E_{\pi,\sigma};T)=\frac{1}{\exp({ \beta E_{\pi,\sigma} })-1}
  \end{equation*}
  with the pion and sigma energies
  \begin{equation*}
    E_\pi = \sqrt{k^2+\overline{\Omega}^{\,\prime}_k}\;~,~ E_\sigma
    =\sqrt{k^2+\overline{\Omega}^{\,\prime}_k+2\rho\,\overline{\Omega}^{\,
        \prime\prime} _k},
  \end{equation*}
where the primes denote derivatives of
  $\overline{\Omega}=\Omega+c\sigma$ with respect to $\rho$ field, $\rho=(\sigma^2+\vec\pi^2)/2$, and $\beta=1/T$.
  The fermion distribution functions $N(\Phi,\Phi^*;T,\theta)$ and
  $\bar{N}(\Phi,\Phi^*;T,\theta)$,
  \begin{eqnarray}\label{n1}
    N(\Phi,\Phi^*;T,\theta)&=&\frac{1+2\Phi^*\exp[\beta E_q- i \theta ]+\Phi \exp[2(\beta E_q- i \theta) ]}{1+3\Phi \exp[2(\beta E_q- i \theta)]+
      3\Phi^*\exp[\beta E_q- i \theta ]+\exp[3(\beta E_q- i \theta)]},  \\
    \bar{N}(\Phi,\Phi^*;T,\theta)&=&N(\Phi^*,\Phi;T,-\theta),
    \label{n2}
  \end{eqnarray}
  are modified because of the  coupling to the gluon field. Finally, the quark energy is given by
  \begin{equation}
    \label{dispertion}
    E_q =\sqrt{k^2+2g^2\rho}.
  \end{equation}
  \end{widetext}

The minimum of the thermodynamic potential is determined by the
stationarity condition
\begin{equation}
  \left. \frac{d \Omega_k}{ d \sigma} \right|_{\sigma=\sigma_k}=\left. \frac{d
      \overline{\Omega}_k}{ d \sigma} \right|_{\sigma=\sigma_k} - c =0.
  \label{eom_sigma}
\end{equation}
The flow equation~(\ref{eq:frg_flow}) is solved numerically with the
initial cutoff $\Lambda=1.2$ GeV (see details in
Ref.~\cite{Skokov:2010wb}).  The initial conditions for the flow are
chosen to reproduce { the  in-vacuum properties:} the physical pion mass $m_{\pi}=138$
MeV, the pion decay constant $f_{\pi}=93$ MeV, the sigma mass
$m_{\sigma}=600$ MeV, and the constituent quark mass $m_q=300$ MeV at
the scale $k\to 0$.  The symmetry breaking term, $c=m_\pi^2 f_\pi$,
corresponds to an external field and consequently does not flow. In
this work, we neglect the flow of the Yukawa coupling $g$, {which is
not expected to be significant for the present studies~(see e.g. Refs.~\cite{Jungnickel,Palhares:2008yq}). }

By solving the equation~(\ref{eq:frg_flow}), one obtains the thermodynamic
potential for the quark and mesonic subsystems, $\Omega_{k\to0} (\Phi,
\Phi^*;T, \theta)$, as a function of the Polyakov loop variables $\Phi$
and $\Phi^*$. The full thermodynamic potential $\Omega(\Phi, \Phi^*;T,
\theta)$ in the PQM model, including quark, meson and gluon
degrees of freedom, is obtained by adding the effective gluon potential ${\cal U}(\Phi,
\Phi^*)$  to $\Omega_{k\to0} (\Phi, \Phi^*;T, \theta)$:
\begin{equation}
  \Omega(\Phi, \Phi^*;T, \theta) = \Omega_{k\to0} (\Phi, \Phi^*;T, \theta) + {\cal U}(\Phi, \Phi^*).
  \label{omega_final}
\end{equation}
At a given temperature and chemical potential, the Polyakov loop
variables, $\Phi$ and $\Phi^*$, are then determined by the stationarity
conditions:
\begin{eqnarray}
  \label{eom_for_PL_l}
  &&\frac{ \partial   }{\partial \Phi} \Omega(\Phi, \Phi^*;T, \theta)  =0, \\
  &&\frac{ \partial   }{\partial \Phi^*}  \Omega(\Phi, \Phi^*;T, \theta)   =0.
  \label{eom_for_PL_ls}
\end{eqnarray}

The thermodynamic potential~(\ref{omega_final}) does not contain
contributions of thermal modes with momenta larger than the cutoff
$\Lambda$.  We take into account the contribution of the high momentum
states by approximating it as quarks interacting only with Polyakov loop degrees of freedom 
as done in Ref.~\cite{Skokov:2010wb}.

\begin{widetext}

\subsection{The mean-field approximation} \label{sec:mf}

\begin{figure*}[t]
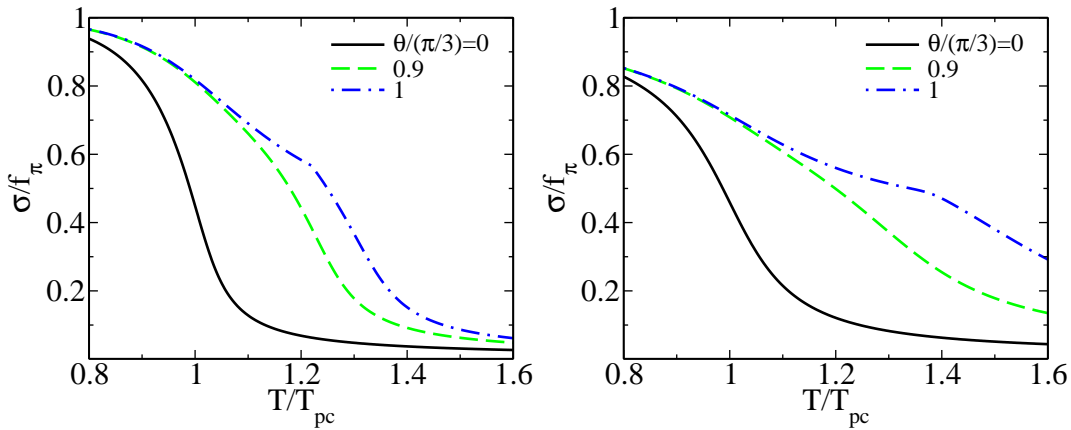

  \includegraphics*[width=7cm]{pol_sig_T_mf}
  \includegraphics*[width=7cm]{pol_sig_T_frg}
  \caption {The chiral order parameter normalized by $f_{\pi}$, as a
    function of temperature for different values of $\theta$ for the {
      PQM} model in the mean-field approximation (left panel) and in
    the FRG approach (right panel).}
  \label{fig:sig_T}
\end{figure*}

\begin{figure*}[t]
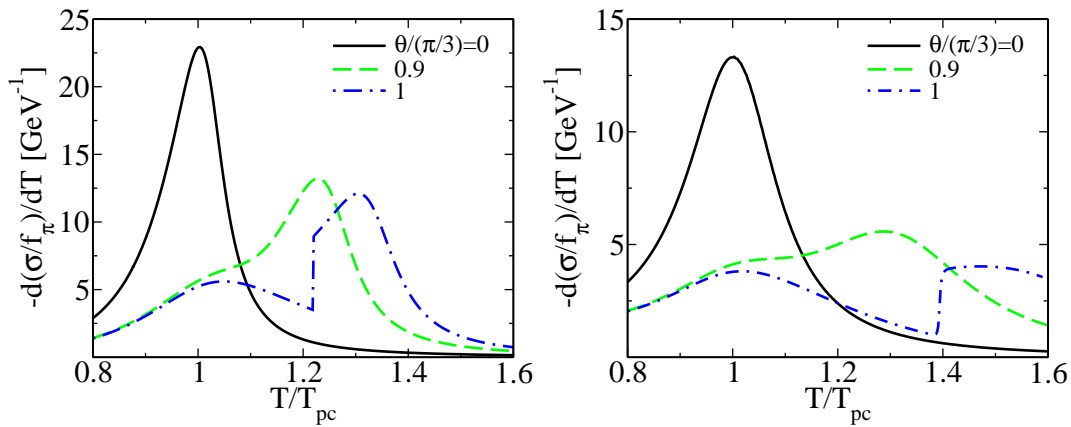

  \includegraphics*[width=7cm]{pol_dsig_dT_T_mf}
  \includegraphics*[width=7cm]{pol_dsig_dT_T_frg}
  \caption {The temperature derivative of the chiral order parameter as a
    function of temperature for different values of $\theta$ for the {
      PQM} model in the mean-field approximation (left panel) and in
    the FRG approach (right panel).}
  \label{fig:dsigdT_T}
\end{figure*}
To show the importance of mesonic fluctuations
on the thermodynamics of the PQM model formulated at imaginary $\mu$ we compare
the FRG results with those obtained in the mean-field
approximation. In the latter, mesonic fluctuations
are neglected and the mesonic fields are replaced by their classical
expectation values.

The thermodynamical potential of the PQM model in the mean-field
approximation reads~\cite{Schaefer:PQM},
\begin{equation}
 \Omega_{\rm MF} = {\cal U}(\Phi,\Phi^*) + U(\langle\sigma\rangle, \langle\pi\rangle=0) + \Omega_{q\bar{q}} (\langle\sigma\rangle,\Phi,\Phi^*).
  \label{Omega_MF}
\end{equation}

Here, the contribution of quarks with the dynamical mass
$m_q=g\langle\sigma\rangle$ is given by
\begin{equation}
    \Omega_{q\bar{q}} (\langle\sigma\rangle, \Phi,\Phi^*) = - \frac{N_c
     N_f}{8 \pi^2} m_q^4 \ln\left(\frac{m_q}{M}\right) - 2 N_f T \int \frac{d^3 p}{(2\pi)^3} \left\{
      \ln
      g^{(+)}(\langle\sigma\rangle, \Phi, \Phi^*; T, \theta) +  \ln
      g^{(-)}(\langle\sigma\rangle,\Phi, \Phi^*; T, \theta) \right\},
    \label{Omega_MF_q}
\end{equation}
where
\begin{eqnarray}
 \label{g}
  g^{(+)}(\langle\sigma\rangle,\Phi, \Phi^*; T, \theta) &=& 1 + 3 \Phi
  \exp[-(\beta E_q-i\theta)] + 3 \Phi^*\exp[-2( \beta E_q-i\theta)] + \exp[-3(\beta E_q-i\theta)], \\
 g^{(-)}(\langle\sigma\rangle,\Phi, \Phi^*; T, \theta) &=& g^{(+)} (\langle\sigma\rangle,\Phi^*, \Phi; T, -\theta)
\end{eqnarray}
\end{widetext}
and $E_q = \sqrt{p^2+m_q^2}$ is the quark quasi-particle energy. The
first term in Eq.~(\ref{Omega_MF_q}) is a vacuum
contribution regularized by dimensional regularization with
renormalization scale $M$ \cite{MFonVT}.
The relevance of the vacuum contribution for the thermodynamics
of chiral models was demonstrated and studied in detail in
Refs.~\cite{MFonVT} and \cite{Nakano:2009ps}.

The equations of motion for the mean fields are obtained by requiring
that the thermodynamic potential is stationary with respect to changes
of $\langle\sigma\rangle$, $\Phi$ and $\Phi^*$. Utilizing the fact that
$\Phi^*$ is the complex conjugate of $\Phi$ at imaginary chemical
potential, we introduce the modulus and the phase of the Polyakov loop
as $\Phi = |\Phi|e^{i\phi}$ and $\Phi^* = |\Phi|e^{-i\phi}$ and the
stationary condition is given as
\begin{equation}
  \frac{\partial \Omega_{\rm MF}}{\partial \langle\sigma\rangle}
   = \frac{\partial \Omega_{\rm MF}}{\partial |\Phi|} = \frac{\partial \Omega_{\rm MF}}{\partial \phi} =0.
  \label{EOM_MF}
\end{equation}
The model parameters are fixed to reproduce the same vacuum physics as
in the FRG calculation.

\section{Thermo\-dynamics of the PQM model}\label{sec:thermo}

\begin{figure*}[t]
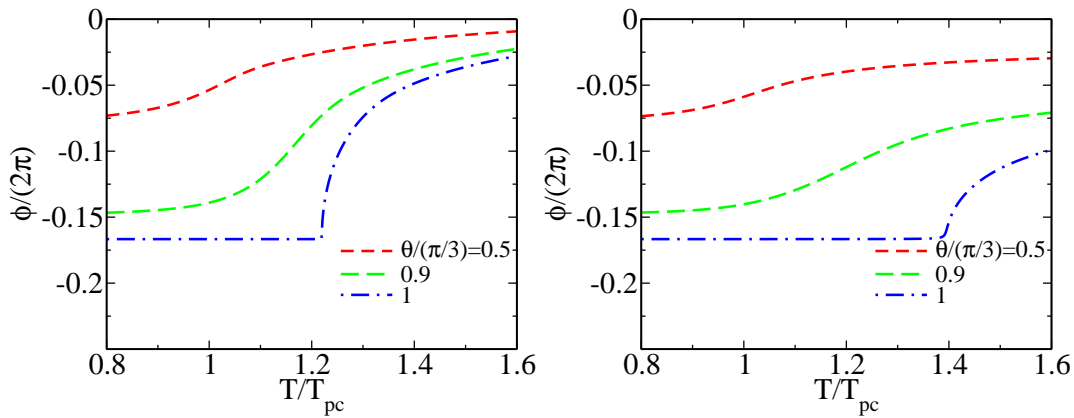

  \includegraphics*[width=7cm]{pol_phi_T_mf}
  \includegraphics*[width=7cm]{pol_phi_T_frg}
  \caption {The phase of the Polyakov loop $\phi$ as a
    function of temperature for different values of $\theta$ for the {
      PQM} model in the mean-field approximation (left panel) and in
    the FRG approach (right panel).}
  \label{fig:phi_T}
\end{figure*}

The PQM model, which is expected to belong to the same universality
class as QCD,  
exhibits 
 a generic phase diagram with a critical point at
a non-vanishing real chemical potential \cite{Schaefer:PQM}. A  detailed comparison between
thermodynamic properties  of the PQM model obtained within the FRG and
in the mean-field approach at real chemical potential was recently
studied in Ref.~\cite{Skokov:2010uh}.

In the following, we explore  the critical properties of the PQM model
at imaginary chemical potential $\mu/T= -i \theta$  using the functional
renormalization group approach and in the mean-field approximation.

Under the mean-field dynamics,  many features of the PQM model at imaginary chemical potential
should be  common with that obtained previously  in the PNJL model. 
Thus, the nature of the phase structure 
in the mean-field approximation for the PNJL model found in
Ref.~\cite{Kenji:2011} also applies to  our  present calculations within
the PQM model. Therefore, we focus on the effects of the mesonic
fluctuations on thermodynamics at imaginary chemical potential and at
finite temperature.

\begin{figure*}[t]
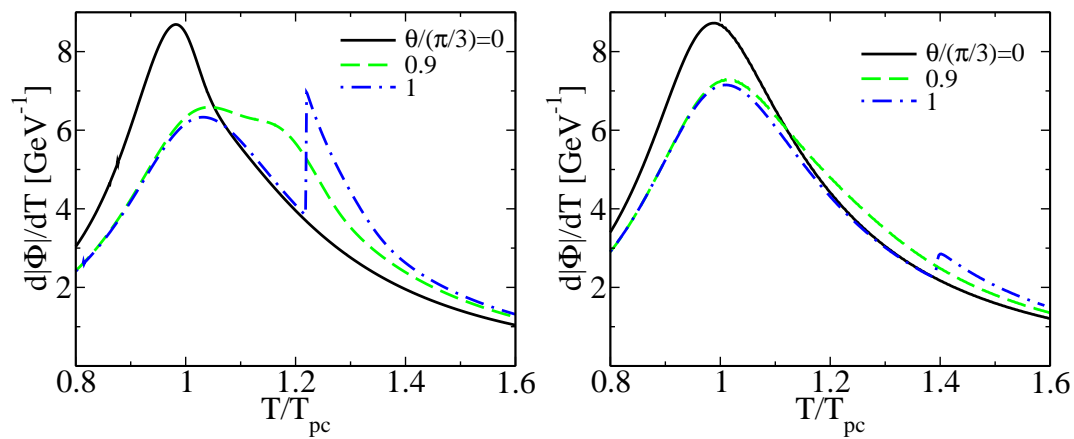

  \includegraphics*[width=7cm]{pol_dmod_dT_T_mf}
  \includegraphics*[width=7cm]{pol_dmod_dT_T_frg}
  \caption {The temperature derivative of modulus of the Polyakov loop  as a
    function of temperature for different values of $\theta$ for the {
      PQM} model in the mean-field approximation (left panel) and in
    the FRG approach (right panel).}
  \label{fig:dmoddT_T}
\end{figure*}

At  imaginary chemical potential, in addition to the chiral transition which is characterized by the chiral order
parameter $\sigma$  and the deconfinement transition which is indicated
by the rapid increase of the modulus of the Polyakov loop, the PQM model
exhibits also  the Roberge-Weiss (RW) transition. 
Since the  thermodynamic potential  of the PQM model shares the
Roberge-Weiss periodicity of QCD, 
$\Omega(T, \theta) = \Omega(T, \theta + \frac{2\pi n}{3})$ with  $n\in N$,
which is a remnant of global $Z(3)$ symmetry of the SU$_c$(3) gauge group,
the first-order RW phase transition takes place in the deconfined phase at $\theta=\pi/3$. The RW transition
can be understood as a sudden change of the phase $\phi$ of the Polyakov
loop. While $\phi$  as a function  of $\theta$ smoothly changes
below the RW endpoint, at $T=T_\text{RW}$ it has a discontinuity
 caused by a transition from one $Z(3)$ sector to
another. This transition also influences the thermodynamic potential and
the modulus of the Polyakov loop as a cusp at $\theta=\pi/3$.
Therefore, at imaginary chemical potential,   the model contains  three variables;  $\sigma$, $|\Phi|$, and
 the phase $\phi$  as the order parameters for  the corresponding phase
transitions.

\begin{figure*}[t]
  \includegraphics*[width=7.5cm]{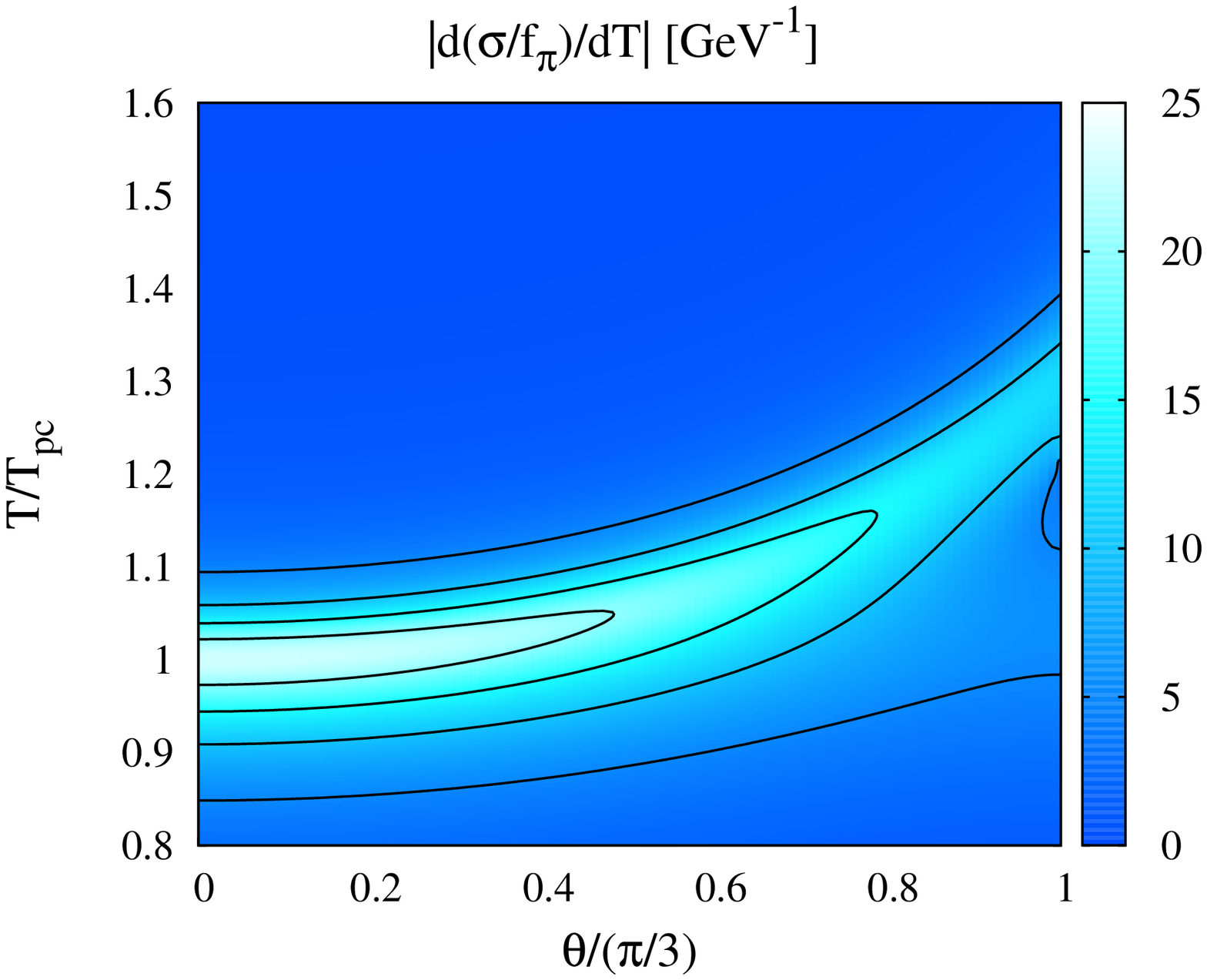}
  \includegraphics*[width=7.5cm]{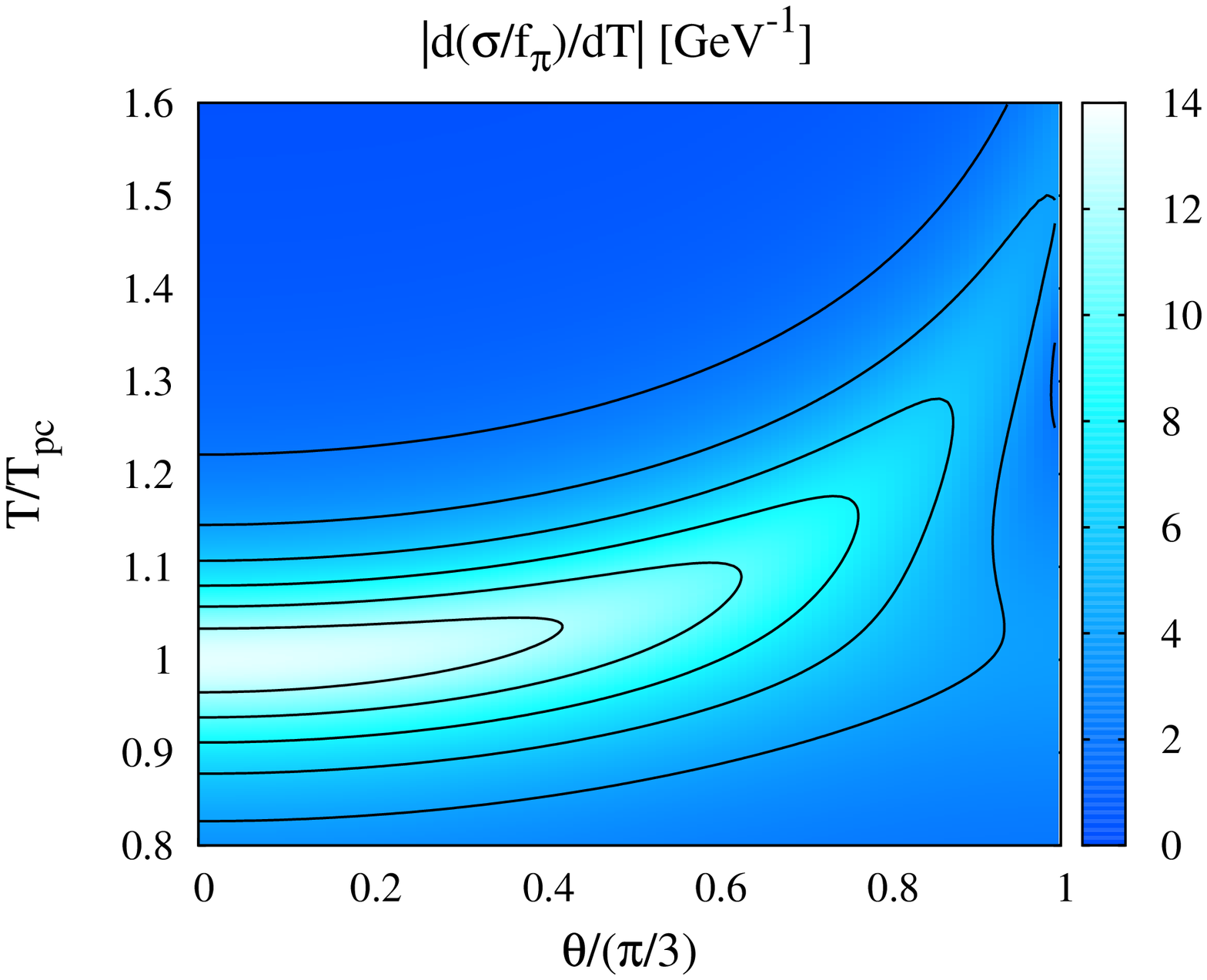}
  \caption {
	  Contour plots of the  temperature  derivative of modulus of the chiral order parameter 
    in the mean-field approximation (left panel)  and in the FRG approach (right panel).}
  \label{fig:dsigmadt2d}
\end{figure*}

\begin{figure*}[t]
  \includegraphics*[width=7.5cm]{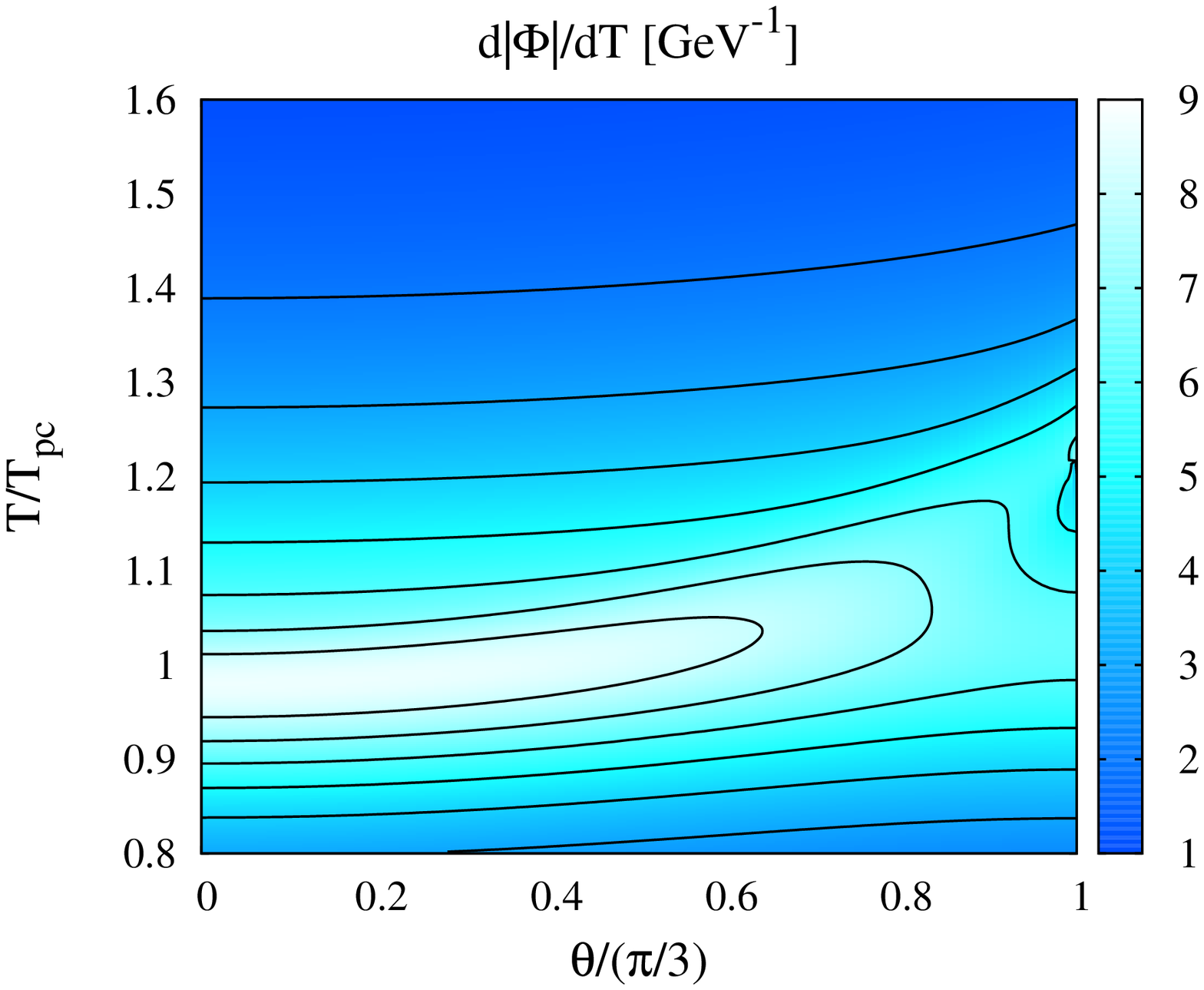}
  \includegraphics*[width=7.5cm]{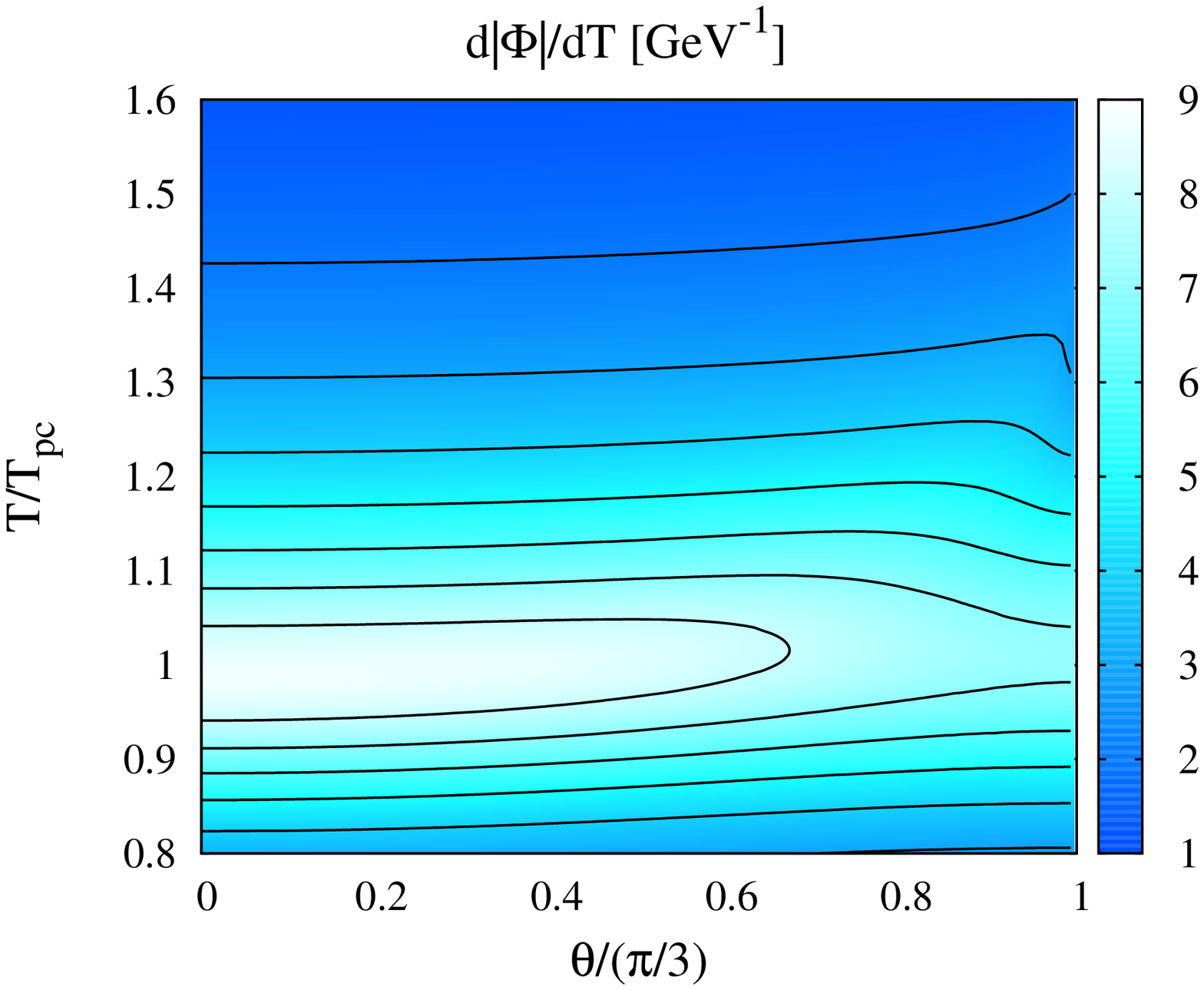}
  \caption {
 Contour plots of the  temperature  derivative	of  the modulus of the Polyakov loop 
    in the mean-field approximation (left panel)  and in the FRG approach (right panel).}
  \label{fig:dldt2d}
\end{figure*}
\subsection{Properties of order parameters in the presence of mesonic fluctuations}

By solving the flow equation (\ref{eq:frg_flow}) for the quantum potential  we can   calculate  properties of all three order parameters discussed above in the presence of mesonic fluctuations.

Figure \ref{fig:sig_T} shows the chiral order parameter normalized by
$f_\pi=93$ MeV as a function of
temperature  scaled by  
the chiral crossover temperature at $\theta=0$, $T_{\text{pc}}$. To identify $T_{\text{pc}}$ the  maximum  of $-d\sigma/dT$ was used. 

 The order parameter is shown in Fig. \ref{fig:sig_T}  for 
various values of the imaginary chemical potential. The FRG results are compared to  that obtained in the mean-field approximation.
There is a clear change in the shape of $\sigma$ with increasing $\theta
$ from 0 to $\theta =\pi/3$. This is the case in the mean-field as well as in
the FRG approach. 
However, the mesonic fluctuations included in the  FRG scheme  
imply a strong  smearing of the order parameter, making  the chiral transition smoother.  
This effect is even more pronounced    in the derivative of the order parameter 
with respect to the temperature.
As seen in Fig.~\ref{fig:dsigdT_T} the 
 peak value of $-d(\sigma/f_\pi)/dT$
in the FRG calculations  at vanishing chemical potential is almost a  half of the
mean field counterpart. From Fig.~\ref{fig:dsigdT_T} one also sees, that in the  mean-field as well as  in the FRG calculations,  the peak
of  $-d(\sigma/f_\pi)/dT$ shifts to higher temperature with increasing  $\theta$.     Applying an  analytical continuation such a shift was used to describe the change of the chiral cross over transition temperature with real chemical potential \cite{deF-P}.

In the present calculation, however, one sees
two peaks structure in $-d(\sigma/f_\pi)/dT$. The ones at higher temperature are
associated with the chiral crossover transition, as it is smoothly
connected from the unique peak at small $\theta$. The first peak, which  is due to the deconfinement transition, was not previously observed in the mean-field calculations in the PNJL model. Its presence in  the PQM model has to be attributed to the specific implementation of the chiral sector.
The discontinuity  which is specific to $\theta=\pi/3$, appears at
$T=1.22~T_{\text{pc}}$ in the mean field and at $T=1.39~T_{\text{pc}}$ in the
FRG calculations. This discontinuity is attributable to the RW endpoint
where the phase $\phi$ of the Polyakov loop shows a critical behavior.  
We note, that in the LGT calculations, the RW end point is located at
$T\approx (1.1-1.2) T_{pc}$~\cite{D'Elia-Lombardo}. The obtained model
results on this temperature strongly depend on the
parametrization of the Polyakov loop potential. For instance, for the
logarithmic potential \cite{Ratti:2007jf}, the discontinuity of
$-d(\sigma/f_\pi)/dT$ at $\theta=\pi/3$ appears at $T=1.08 T_{pc}$ for
the FRG approach and $T=1.09T_{pc}$ in the mean-field approximation.

Figure \ref{fig:phi_T} shows the  $\phi$ as a function of temperature.
\footnote{$\theta=\pi/3$ above $T_{\text{RW}}$ is the first-order
transition line where two minima coexist and $T_{\text{RW}}$ is the
bifurcation. For simplicity in the
figure, we omit the other minimum in $\phi$.}
We find in the PQM model with the polynomial potential
\eqref{eff_potential} the second-order transition both in the FRG approach and in the mean-field approximation.
Thus, the RW endpoint is a critical point  located at $\theta =
\pi/3 + 2\pi n/3 $ and at  $T_{\rm RW}/T_{\rm pc} = 1.39$ in the FRG
approach and  at a lower temperature    $T_{\rm RW}/T_{\rm pc} = 1.22$ in
the mean-field approximation. Although the order of the transition
remains unchanged by the fluctuations, in the FRG calculations the $\phi(T)$ 
is a smoother function of temperature.  
The Polyakov loop is treated within the  mean-field approximation also in 
the FRG approach, thus the observed  smoothening   of $\phi(T)$  in Fig. \ref{fig:phi_T} (right)  appears  owing to its coupling to the chiral order parameter which is
 smeared by the quantum fluctuations.  

The effect of the fluctuations on the deconfinement transition is   more transparent if  
the
derivative of  $|\Phi|$ with respect to the
temperature is considered.
The  $d|\Phi|/dT$ as a function of $T$ is  depicted in Fig.~\ref{fig:dmoddT_T} in the FRG and mean-field calculations. 
 At small  $\theta$  
 this derivative  has a unique
peak, that is attributed to the location of the
deconfinement transition. However, 
 near $\theta=\pi/3$, where the phase $\phi$ also
changes rapidly,  an additional   peak appears. In the mean-field calculations   one sees
two well-separated  maxima.
At $\theta=\pi/3$, where the second-order RW endpoint appears 
as the rapid change of $\phi$ seen in  Fig.~\ref{fig:phi_T}, there is 
a discontinuity in  $d|\Phi|/dT$ at $T=T_{\text{RW}}$, both   in the mean-field and in the FRG calculations. However,
 the jump  associated with the RW endpoint is smaller in the FRG case.
This, 
as shown in Figs. \ref{fig:sig_T} and \ref{fig:phi_T}, is due to  mesonic fluctuations which  smoothen the
critical behavior of the chiral order
parameter and the phase $\phi$. Consequently, as seen in 
Fig.~\ref{fig:dmoddT_T} (right), the discontinuity of $d|\Phi|/dT$ at
$\theta=\pi/3$ induced by the RW endpoint becomes  weaker.   
Nevertheless, in these studies,  the deconfinement transition,
defined by the peak of  $d|\Phi|/dT$, is now separated from the RW
transition contrary to previous analysis in the  PNJL model~\cite{Kyushu_PNJL,Kenji:2011}.
One should note, however, that such  separation occurs only in the case of the
polynomial Polyakov loop potential \eqref{eff_potential}, which violates
the restriction on the Polyakov loop target space required by the $SU(3)$
Haar measure \cite{Kenji:2011}. If one uses the logarithmic potential
\cite{Ratti:2007jf} which exhibits stronger transition, the deconfinement
critical line connects to the RW endpoint, as expected.





\subsection{Phase diagrams}

\begin{figure*}[t]
  \includegraphics*[width=7.5cm]{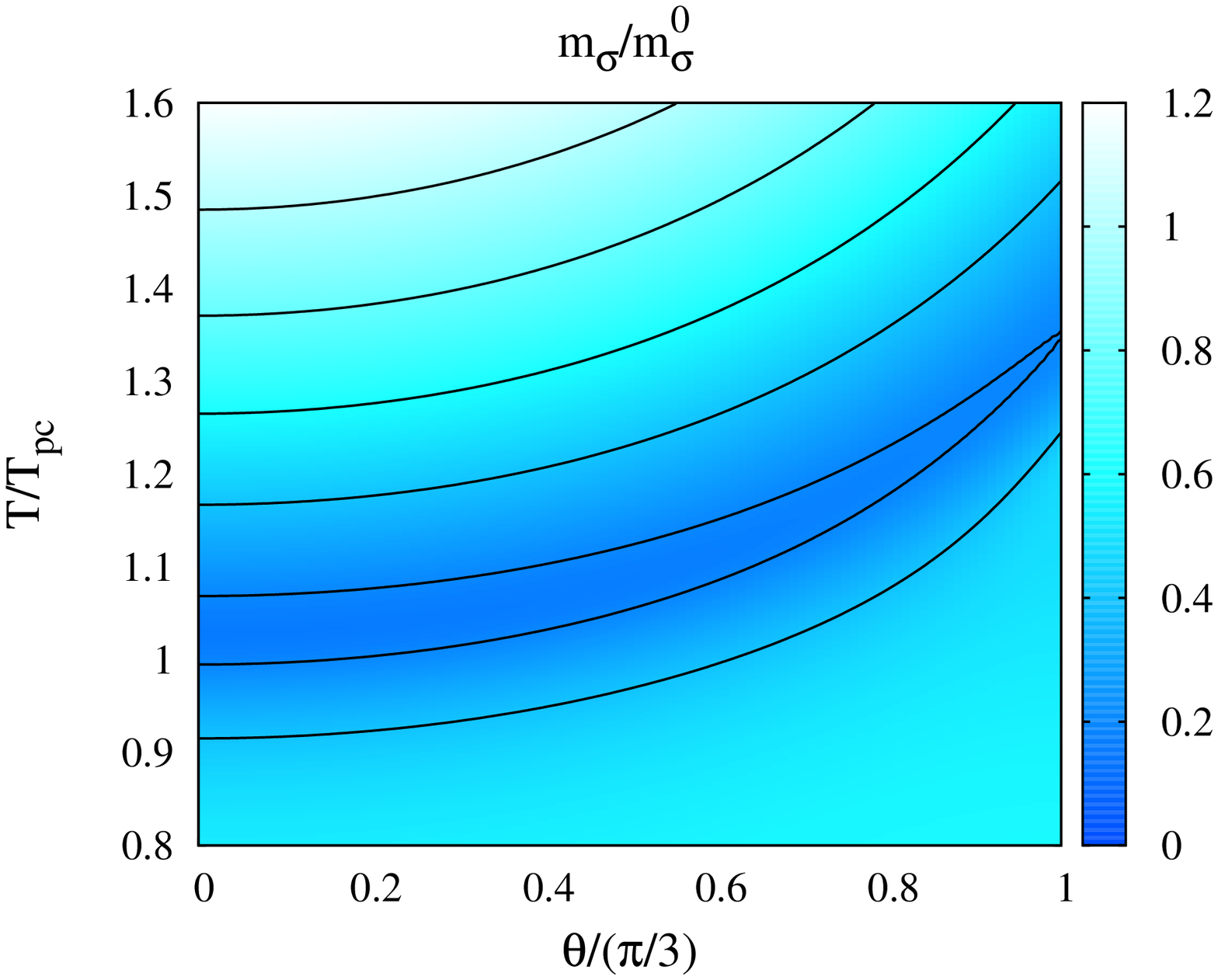}
  \includegraphics*[width=7.5cm]{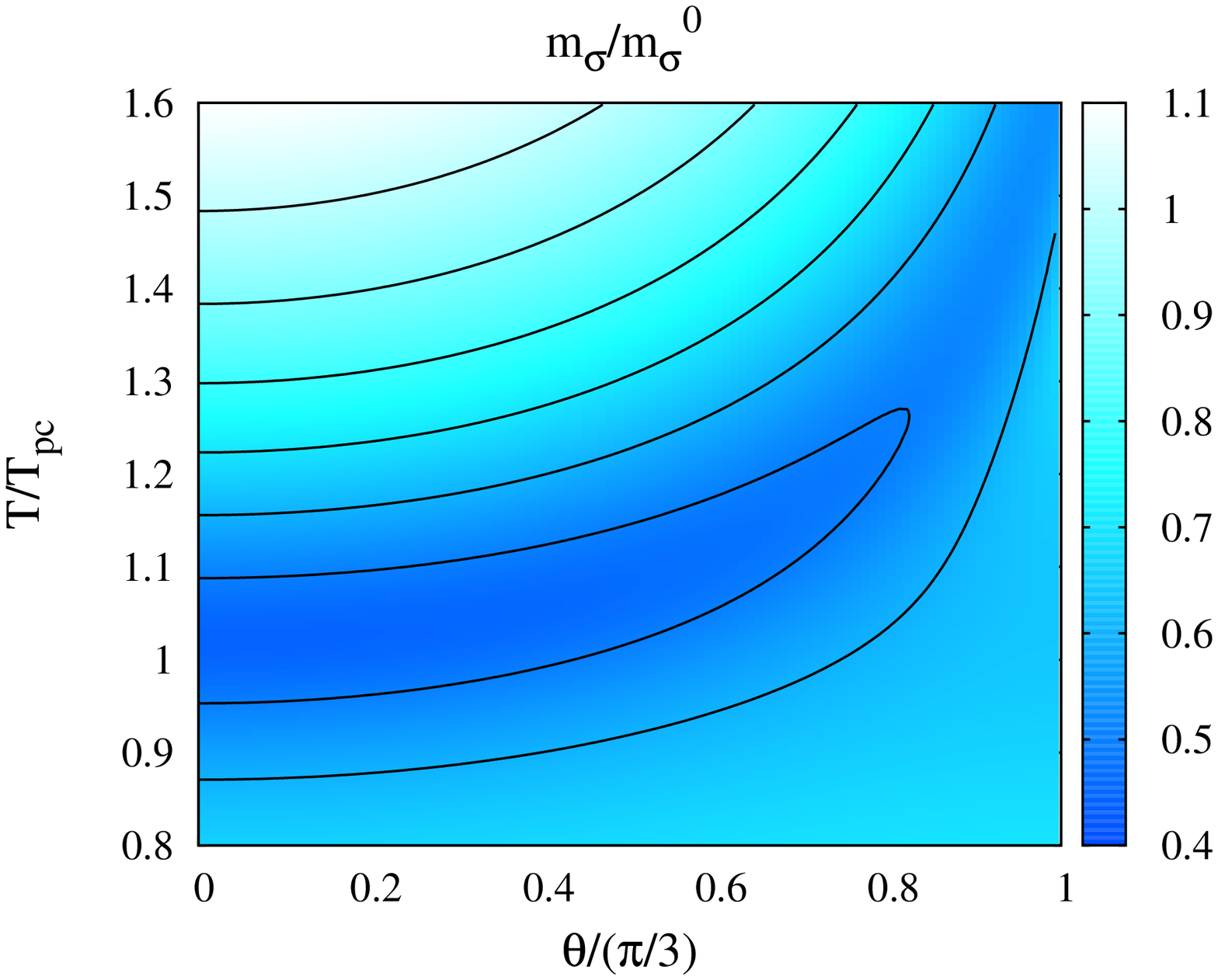}
  \caption {
 Contour plots of the   sigma mass
    in the mean-field approximation (left panel)  and in the FRG approach (right panel).}
  \label{fig:msigma2d}
\end{figure*}

The properties  of order parameters in the PQM model discussed above,  can  be also represented as the 
phase diagram  in the  $(T-\theta)$ plane. For  illustration, we show in Figs.~\ref{fig:dsigmadt2d}  and ~\ref{fig:dldt2d} contour
maps of   
$-d(\sigma/f_\pi)/dT$ for  the chiral transition and  $d|\Phi|/dT$ for the
deconfinement.
The effect of fluctuations can be clearly seen in the changes of the
contours.
The crossover chiral line, defined by the maximum of $-d(\sigma/f_\pi)/dT$,
meets the first-order RW transition at $T_{\rm ce}/T_{\rm pc} = 1.47$ if
the fluctuations are included, and at $T_{\rm ce}/T_{\rm pc} = 1.30$
in the mean-field approximation.
However, as discussed above, at imaginary chemical potential, the transition points for the chiral
and deconfinement crossover  are not determined unambiguously
 as they depend on the particular choice  of the Polyakov loop potential. 
 In the present calculations,  the double peak
structure can be  read off from the contour maps.

 An alternative
determination of the chiral critical line is based on the location of the minimum in  the sigma meson
mass $m_\sigma$.  Figure~\ref{fig:msigma2d} shows  the contour maps for $m_\sigma$, normalized to its vacuum value   $m_\sigma^0$ . One sees in this figure that the double peak
structure induced by the deconfinement does not show up  in the sigma mass.
The effect of  fluctuations  appears as a smearing of the sharp
minimum and the stronger curvature. This is evident in Fig.~\ref{fig:msigma2d} when comparing results  obtained in   the mean-field approximation (left) with the FRG approach (right).


\section{Summary and Conclusions}\label{sec:concl}
We have formulated and explored the thermodynamics of the Polyakov loop extended
quark--meson model (PQM) at imaginary chemical potential, including mesonic
fluctuations within the functional renormalization group method
(FRG). The flow equations for the scale-dependent
thermodynamic potential at finite temperature and at imaginary chemical potential were solved in the
presence of a background gluonic field.

We have shown that the non-perturbative fluctuations included in the FRG
approach have an important effect on the critical properties of the
system. Specifically, the fluctuations smoothen the chiral phase transition.
We find that this leads to modification of the curvature of the chiral critical
line. We also find that the Polyakov loop variables (modulus and phase)
are also smeared by fluctuations of the chiral order parameter.
We point out that the deconfinement crossover transition and the 
second-order Roberge-Weiss endpoint can be separated if the polynomial
Polyakov loop effective potential is used to model the gluon thermodynamics. 

Our results show, that fluctuations of the meson field and
non-perturbative effects have to be included when studying critical
behavior of  effective chiral models. Such fluctuations can
modify substantially the model properties  and predictions. This is
particularly important when one attempts to quantify lattice QCD
results within effective chiral models.

\section*{Acknowledgments}
V. Skokov acknowledges the support by the Frankfurt Institute for
Advanced Studies (FIAS). K. Morita is supported by the Yukawa International
Program for Quark-Hadron Sciences (YIPQS) at Kyoto University.
 K. Redlich acknowledges partial support by  the Polish
Ministry of Science.


\end{document}